\documentclass[a4paper,12pt]{article}
\usepackage{graphicx}
\usepackage{comment}
\usepackage{hyperref}
\usepackage{makecell}
\RequirePackage{color}
\RequirePackage{mathtools}
\RequirePackage{snapshot}
\RequirePackage[english]{babel}
\RequirePackage[utf8]{inputenc}
\RequirePackage{cuted}
\RequirePackage{booktabs}
\RequirePackage{multirow}
\RequirePackage{url}
\RequirePackage[numbers,sort&compress]{natbib}
\RequirePackage[normalem]{ulem}
\RequirePackage{changepage}
\usepackage{caption}
\newcommand\pubnumber{TTK-22-42, P3H-22-115}
\newcommand\pubdate{\today}

\def\institute{Institute for Theoretical Particle Physics
and Cosmology, RWTH Aachen University, \\D-52056 Aachen, Germany}
\def\support{\footnote{Work supported by German Research Foundation (DFG) Collaborative Research Centre/Transregio project
CRC/TRR 257:P3H - \textit{Particle Physics Phenomenology after the Higgs Discovery}, by the Research Training
Group GRK 2497: \textit{The physics of the heaviest particles at the Large Hadron Collider} and by a grant of the
Bundesministerium für Bildung und Forschung (BMBF).}}

\def\Title#1{\begin{center} {\Large #1 } \end{center}}
\def\Author#1{\begin{center}{ \sc #1} \end{center}}
\def\Address#1{\begin{center}{ \it #1} \end{center}}

\newcommand\pubblock{\rightline{\begin{tabular}{l} \pubnumber\\
         \pubdate  \end{tabular}}}
\newenvironment{Abstract}{\begin{quotation}  }{\end{quotation}}
\newenvironment{Presented}{\begin{quotation} \begin{center} 
             PRESENTED AT\end{center}\bigskip 
      \begin{center}\begin{large}}{\end{large}\end{center} \end{quotation}}

\def\beq{\begin{equation}}
\def\eeq#1{\label{#1}\end{equation}}
\def\eeqn{\end{equation}}

\def\beqa{\begin{eqnarray}}
\def\eeqa#1{\label{#1}\end{eqnarray}}
\def\eeqan{\end{eqnarray}}

\let\bar=\overbar

\def\Dslash{\not{\hbox{\kern-4pt $D$}}}
\def\dslash{\not{\hbox{\kern-2pt $\del$}}}

\def\msb{{\bar{\ssstyle M \kern -1pt S}}}

\begin{document}
\begin{titlepage}
\pubblock

\vfill
\Title{NLO QCD corrections and off-shell effects for $t\bar{t}H$ production in the Higgs characterisation model}
\vfill
\Author{ Jonathan Hermann\support}
\Address{\institute}
\vfill
\begin{Abstract}
In these proceedings we discuss $t\bar{t}H$ production at the Large Hadron Collider in the Higgs characterisation model. We demonstrate the importance of off-shell effects and higher-order QCD corrections for this process and highlight the importance of single-resonant contributions for the production of a $\mathcal{CP}$-odd Higgs boson.

\end{Abstract}
\vfill
\begin{Presented}
$15^\mathrm{th}$ International Workshop on Top Quark Physics\\
Durham, UK, 4--9 September, 2022
\end{Presented}
\vfill
\end{titlepage}
\def\thefootnote{\fnsymbol{footnote}}
\setcounter{footnote}{0}

\section{Introduction}

Despite its small cross-section compared to gluon-gluon and vector-boson fusion, $t\bar{t}H$ production is one of the most important Higgs production channels at the Large Hadron Collider (LHC). It is particularly interesting for probing the top-Higgs Yukawa interaction as it already appears at tree level in $t\bar{t}H$ production. The $\mathcal{CP}$-nature of this interaction has been of great interest since the discovery of the Higgs boson in 2012. Both CMS and ATLAS have performed measurements of the mixing angle $\alpha_{CP}$ between a 
$\mathcal{CP}$-even and a $\mathcal{CP}$-odd Higgs boson \cite{CMS:CP, ATLAS:CP}. In both cases the results are in agreement with the Standard Model (SM) prediction of a $\mathcal{CP}$-even Higgs boson and the measurements have allowed the exclusion of a pure $\mathcal{CP}$-odd state with $3.7 \,\sigma$ by CMS and $3.9 \,\sigma$ by ATLAS. Nonetheless, there still remains a lot of freedom in the mixing angle and CMS has actually found that their fit results favour a $\mathcal{CP}$-mixed coupling.

With our work we aim to provide state-of-the-art predictions for $t\bar{t}H$ production in the Higgs characterisation model \cite{Artoisenet} which extends the SM Lagrangian by a $\mathcal{CP}$-odd top-Higgs Yukawa interaction term. To this end, we present results for this process in the dilepton decay channel at next-to-leading order (NLO) in QCD and including full off-shell effects, i.e. $ p p \to b \bar{b} e^+ \mu^- \nu_e \bar{\nu}_\mu H + X$ production at order $\mathcal{O} \left( \alpha_s^3 \alpha^5 \right)$ for the LHC at $\sqrt{s}=13$ TeV. Here, full off-shell effects mean that we describe the unstable intermediate particles in the complex mass scheme and include all double-, single- and non-resonant contributions. These results are based on the ones presented in Ref. \cite{Hermann} and were computed using the HELAC-NLO framework \cite{HELAC1, HELAC2}. 
Similar calculations have already been performed for this process in the SM case in Refs. \cite{Denner:2015yca, Denner:2016wet, Stremmer} but not for $\mathcal{CP}$-mixed top-Higgs interactions. For the latter, the previous state-of-the-art predictions only included NLO QCD corrections to the $t\bar{t}H$ production and not to the top-quark decays. In addition, only double-resonant diagrams were considered \cite{Demartin}.

\section{Phenomenological results}
\begin{table*}[t!]
    \caption{ \it Integrated fiducial cross-sections calculated  in the NWA, NWA with LO top-quark decays and full off-shell approach for $\alpha_{CP} = 0, \pi/4$ and $\pi/2$. Table was taken from \cite{Hermann}. }
    \label{table:integrated}
    
    \centering
    \renewcommand{\arraystretch}{1.5}
    \begin{tabular}{l@{\hskip 3mm}l@{\hskip 3mm}ll@{\hskip 3mm}l}
        \hline\noalign{\smallskip}
        $\alpha_{CP}$ &   & Off-shell & NWA & \makecell{Off-shell \\ effects } \\
        \noalign{\smallskip}\midrule[0.5mm]\noalign{\smallskip}
        \multirow{3}{*}{$0$ (SM)}     & $\sigma_{\text{LO}}$ [fb] & $ 2.0313(2)^{+0.6275\,(31\%)}_{-0.4471 \,(22\%)} $ & $ 2.0388(2)^{+0.6290 \, (31\%)}_{-0.4483 \, (22\%)}$ & $ -0.37\%$\\
        & $\sigma_{\text{NLO}}$ [fb] & $ 2.466(2)^{+0.027 \, (1.1\%)}_{-0.112 \, (4.5\%)} $ & $ 2.475(1)^{+0.027 \, (1.1\%)}_{-0.113 \, (4.6\%)} $ & $ -0.36\%$\\
        & $\sigma_{\text{NLO}_{\text{LOdec}}}$ [fb] & $ - $ & $ 2.592(1)^{+0.161 \, (6.2\%)}_{-0.242 \, (9.3\%)} $ & \\
        \noalign{\smallskip}\hline\noalign{\smallskip}
        &  ${\cal K} = \sigma_{\text{NLO}} / \sigma_{\text{LO}}$    & $ 1.21 $ & $ 1.21 $ (LOdec: $ 1.27$) & \\
        
        \noalign{\smallskip}\midrule[0.5mm]\noalign{\smallskip}
        \multirow{3}{*}{$\pi / 4$}   & $\sigma_{\text{LO}}$ [fb] & $ 1.1930(2)^{+0.3742 \, (31\%)}_{-0.2656 \, (22\%)} $ & $ 1.1851(1)^{+0.3707 \, (31\%)}_{-0.2633 \, (22\%)} $ & $ 0.66\%$\\
        &  $\sigma_{\text{NLO}}$ [fb] & $ 1.465(2)^{+0.016 \, (1.1\%)}_{-0.071 \, (4.8\%)} $ & $ 1.452(1)^{+0.015 \, (1.0\%)}_{-0.069 \, (4.8\%)} $ & $ 0.89 \%$\\
        &  $\sigma_{\text{NLO}_{\text{LOdec}}}$ [fb] & $ - $ & $  1.517(1)^{+0.097 \, (6.4\%)}_{-0.144 \, (9.5\%)} $ & \\
        \noalign{\smallskip}\hline\noalign{\smallskip}
        &  ${\cal K} = \sigma_{\text{NLO}} / \sigma_{\text{LO}}$    & $ 1.23 $ & $ 1.23 $ (LOdec: $ 1.28$) & \\
        
        \noalign{\smallskip}\midrule[0.5mm]\noalign{\smallskip}
        \multirow{3}{*}{$\pi / 2$}   & $\sigma_{\text{LO}}$ [fb] & $ 0.38277(6)^{+0.13123\, (34\%)}_{-0.09121\, (24\%)} $ & $ 0.33148(3)^{+0.11240\, (34\%)}_{-0.07835 \, (24\%)} $ & $ 13.4\%$\\
        &   $\sigma_{\text{NLO}}$ [fb] & $ 0.5018(3)^{+0.0083 \, (1.2\%)}_{-0.0337 \, (6.7\%)} $ & $ 0.4301(2)^{+0.0035 \, (0.8\%)}_{-0.0264 \, (6.1\%)} $ & $ 14.3 \%$\\
        &   $\sigma_{\text{NLO}_{\text{LOdec}}}$ [fb] & $ - $ & $ 0.4433(2)^{+0.0323 \, (7.3\%)}_{-0.0470 \, (11\%)} $ & \\
        \noalign{\smallskip}\hline\noalign{\smallskip}
        &  ${\cal K} = \sigma_{\text{NLO}} / \sigma_{\text{LO}}$    & $ 1.31 $ & $ 1.30 $ (LOdec: $ 1.34$) & \\
        \noalign{\smallskip}\hline\noalign{\smallskip}
    \end{tabular}
\end{table*}

First we compare the integrated fiducial cross-sections for the $\mathcal{CP}$-even, -mixed and -odd Higgs boson production in the narrow-width approximation (NWA) and the full off-shell treatment at LO and NLO in QCD. The results are listed in Table \ref{table:integrated}. 
We find that the cross-sections for the $\mathcal{CP}$-mixed and $\mathcal{CP}$-even cases are about $3$ and $5$ times larger than for the $\mathcal{CP}$-odd one. NLO corrections are around $20 \%$ for the $\mathcal{CP}$-even and -mixed and about $30 \%$ for the $\mathcal{CP}$-odd Higgs boson. In all cases these corrections are consistent between the NWA and the full off-shell treatment and within the LO uncertainties.  Concerning the off-shell effects, we find that these are of the expected order of $\Gamma_t / m_t \sim 0.8 \%$ for the $\mathcal{CP}$-even and -mixed cases and thus negligible at the level of integrated fiducial cross-sections when compared to the NLO scale uncertainties of around $5 \%$. However, in the pure $\mathcal{CP}$-odd case, the off-shell effects are significantly larger than these uncertainties at $14 \%$ for the NLO results and should thus be taken into account even for the integrated fiducial cross-section.

\begin{figure}[t!]
	\begin{center}
		\includegraphics[width=0.35\textwidth]{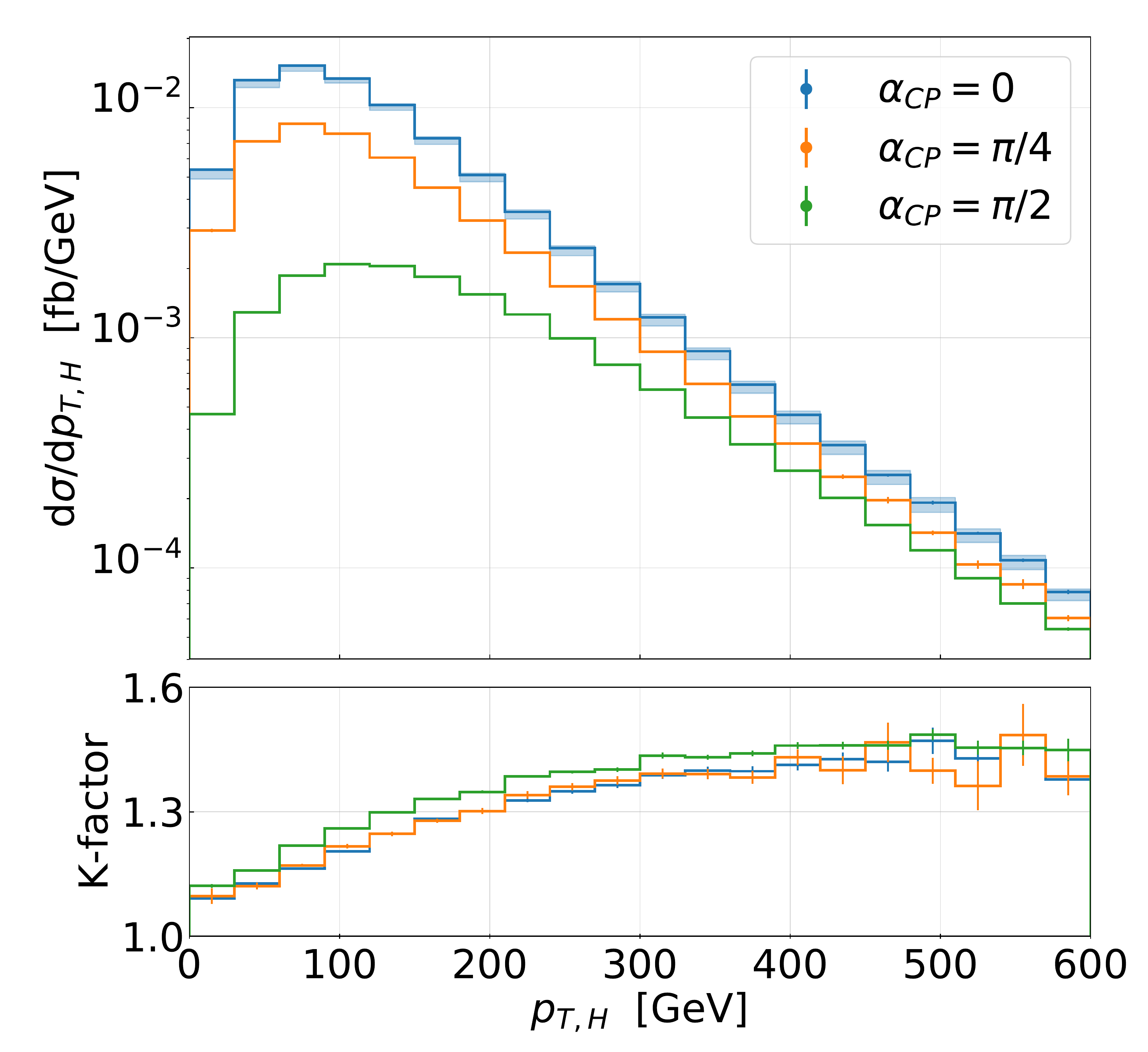}
		\includegraphics[width=0.35\textwidth]{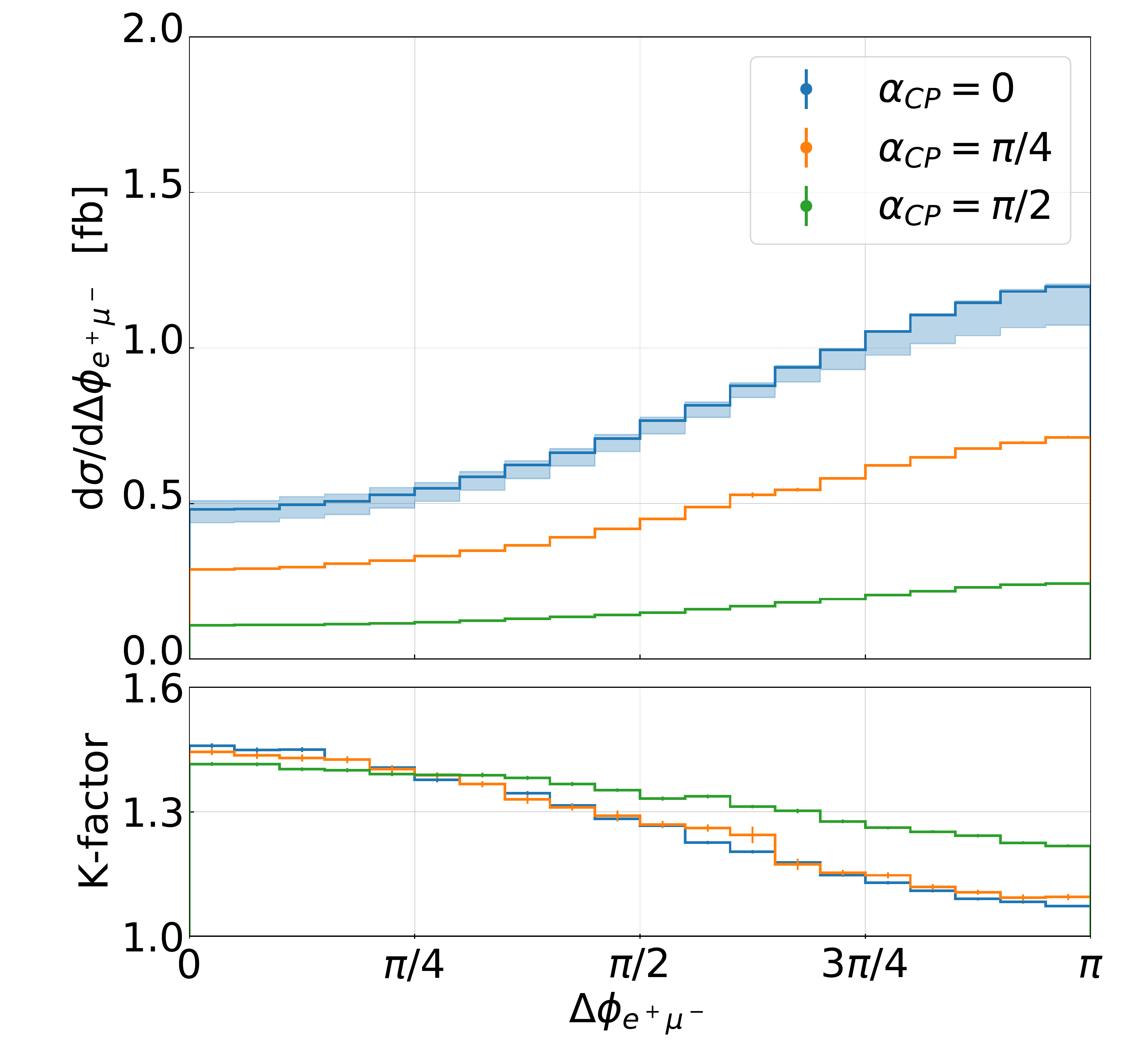}
	\end{center}
	\caption{\label{fig:kfac1a}   \it
		Differential distributions for the observables $p_{T,\,H}$ and $\Delta \phi_{e^+ \mu^-}$ at NLO in QCD. The lower panels show the differential $\mathcal{K}\textrm{-factors}$. Figures were taken from \cite{Hermann}.}
\end{figure}
\begin{figure}[t!]
	\begin{center}
		\includegraphics[width=0.32\textwidth]{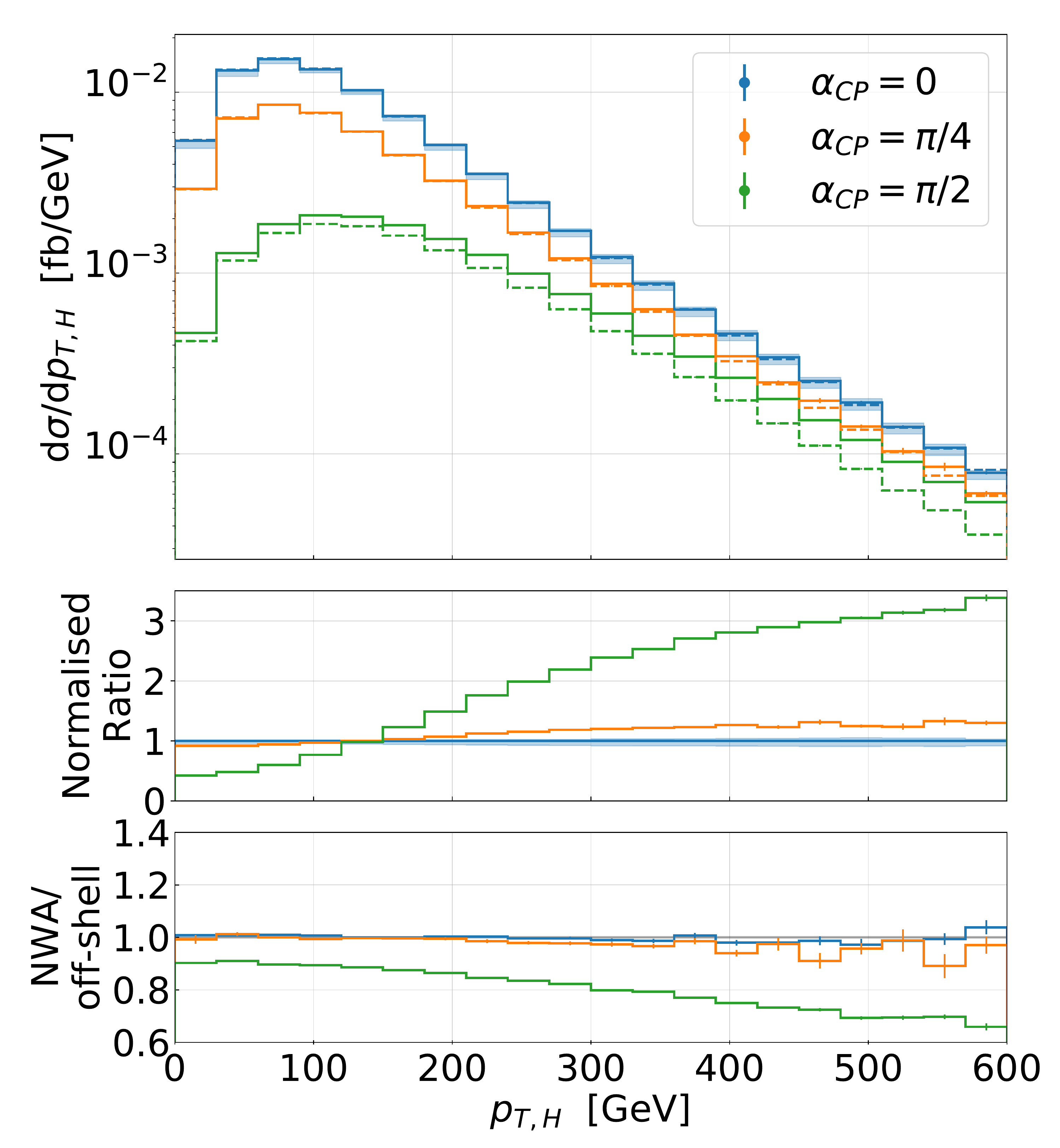}
        \includegraphics[width=0.32\textwidth]{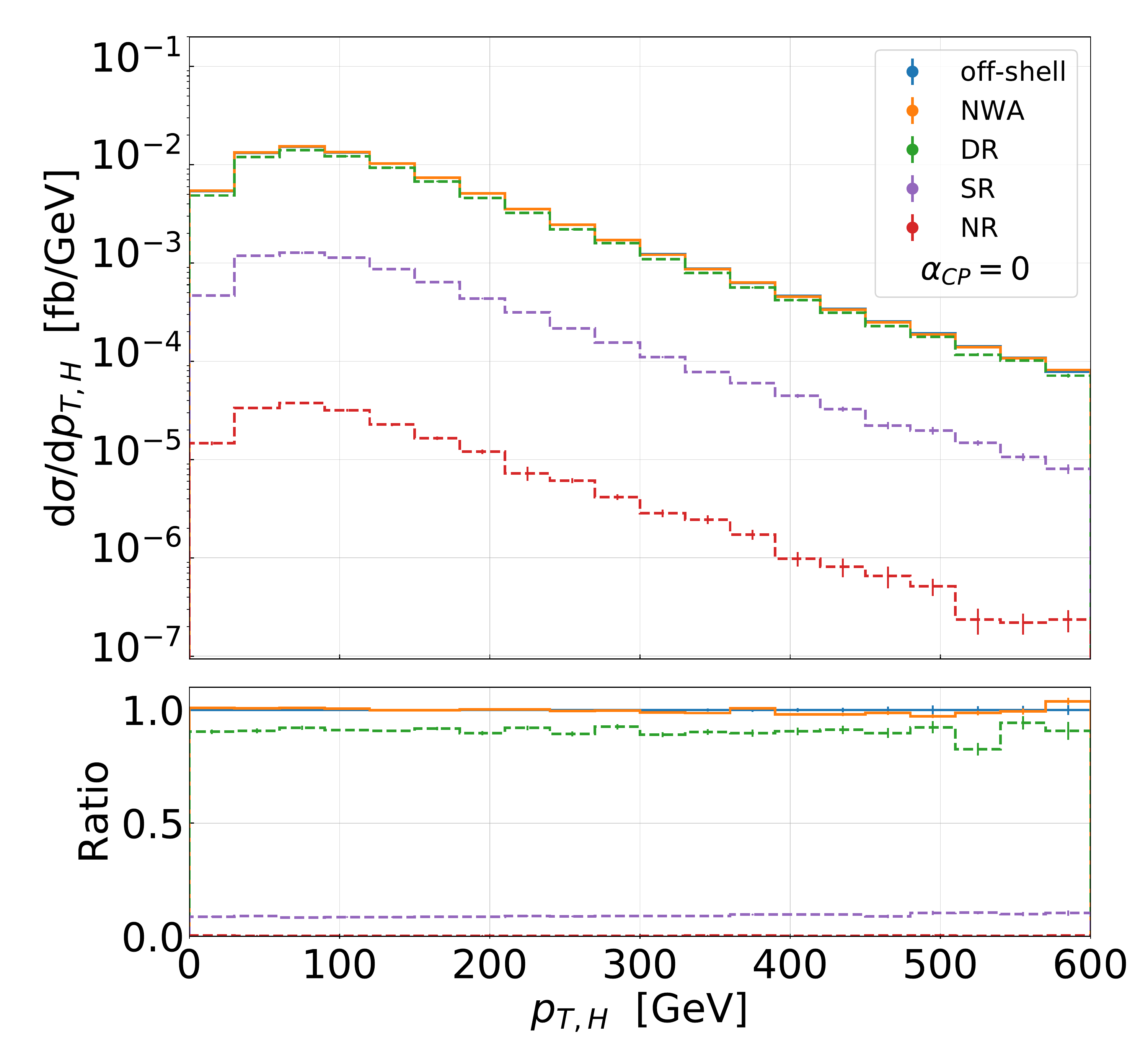}
        \includegraphics[width=0.32\textwidth]{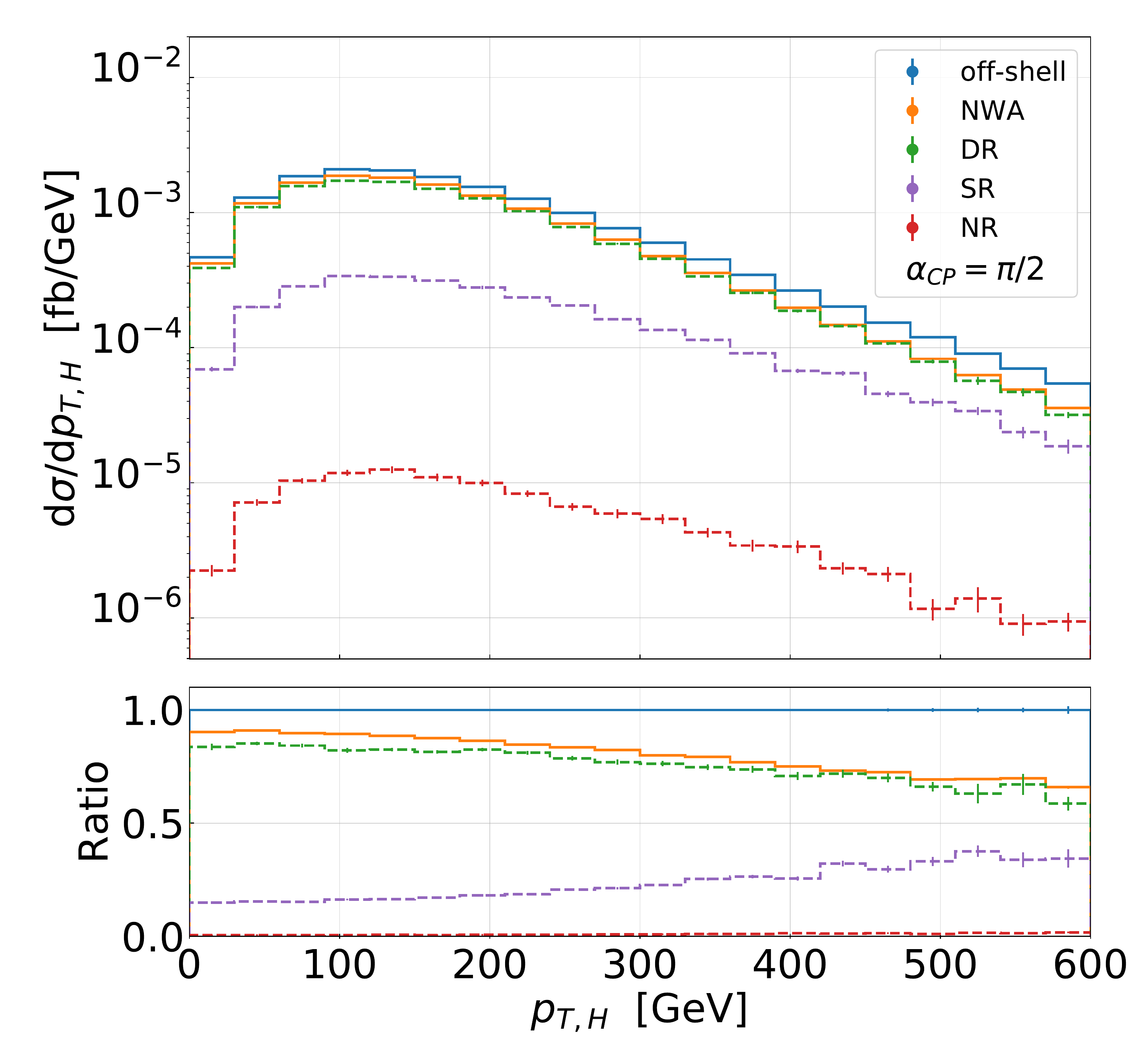}
	\end{center}
	\caption{\label{fig:cosa1} \it
		Left: Differential distributions at NLO in QCD for $p_{T,H}$ for the full off-shell case (solid lines) and the NWA (dashed lines). The ratio to $\alpha_{CP}=0$ of the normalised differential distributions for the full off-shell case is shown in the middle panel, the ratio NWA/off-shell is given in the lower one. \\
        Center: Differential distributions at NLO in QCD for $p_{T,H}$ for the full off-shell case and the NWA as well as the double- (DR), single- (SR) and non-resonant (NR) contributions for the SM case. \\
        Right: Same as central plot but for the $\mathcal{CP}$-odd case. 
        Figures were taken from \cite{Hermann}.} 
\end{figure}

The general trend of larger corrections for the $\mathcal{CP}$-odd Higgs boson can also be observed at the differential level. In Figure \ref{fig:kfac1a}, where we show the differential distributions for $p_{T,H}$ and $\Delta \phi_{e^+,\mu^-}$, one can for example see that NLO corrections are generally larger in the $\mathcal{CP}$-odd case than for the others. For most observables the $\mathcal{K}$-factors follow the same trend as for $p_{T,H}$, i.e. the corrections increase towards the distribution tail but the behaviour of the three $\mathcal{K}$-factors is roughly the same. However, for observables which involve decay products of both top quarks, like $\Delta \phi_{e^+,\mu^-}$, we observe that the $\mathcal{K}$-factor for the $\mathcal{CP}$-odd Higgs boson is flatter than for the other two cases. This happens due to the harder Higgs boson radiation for the former case which suppresses the real radiation corrections that are responsible for the large $\mathcal{K}$-factors for small opening angles and large transverse momenta.

Off-shell effects are also enhanced at the differential level, as can be seen from the lower panel of the left plot in Figure \ref{fig:cosa1}. For the depicted $p_{T,H}$ distribution, they exceed $30 \%$ in the high-$p_{T}$ region for the $\mathcal{CP}$-odd Higgs boson. On the other hand, the effects remain negligible for the other two considered $\mathcal{CP}$-states.
The different size of the off-shell effects can be attributed to the single-resonant (SR) contributions included in the full off-shell computation. In the central and right plot in Figure \ref{fig:cosa1} we show the splitting of the off-shell distribution into its double-, single- and non-resonant parts as defined in Eq. (5.6) to (5.9) Ref. \cite{Hermann}. In the SM case (central plot), the ratio between the three contributions remains virtually constant throughout the entire spectrum which mimics the behaviour of the off-shell effects in this case. In contrast, for the $\mathcal{CP}$-odd state (right plot), the DR contribution decreases significantly towards the distribution tail, in line with NWA distribution. At the same time, the SR contribution increases to account for the difference. Hence, the larger off-shell effects in the latter case can be explained by the different behaviour of the SR contribution.

\section{Summary}
In these proceedings we have presented integrated and differential fiducial cross-sections for $t\bar{t}H$ production with leptonic top-quarks decays in the Higgs characterisation framework.
We have underlined the importance of both off-shell effects and higher-order QCD corrections for these predictions, particularly for the production of a $\mathcal{CP}$-odd Higgs boson. In addition, we have demonstrated that the large off-shell effects in the $\mathcal{CP}$-odd case are a result of the missing single-resonant contributions in the NWA.


\begin{thebibliography}{99}


\bibitem{CMS:CP}
 CMS Collaboration,
\href{https://doi.org/10.48550/arXiv.2208.02686}{{\ttfamily arXiv:2208.02686}}.
%
\bibitem{ATLAS:CP}
 ATLAS Collaboration,
\href{ 	
https://doi.org/10.1103/PhysRevLett.125.061802}{\emph{Phys. Rev. Lett.} \textbf{125} (2020) 061802}.
%
\bibitem{Artoisenet}
P.~Artoisenet \textit{et al.},
  \href{https://doi.org/10.1007/JHEP11\%282013\%29043}{\emph{JHEP} {\bfseries 11} (2013) 043}.
%
\bibitem{Hermann}
J.~Hermann, D.~Stremmer, M.~Worek,
\href{https://doi.org/10.1007/JHEP09(2022)138}{\emph{JHEP} \textbf{09} (2022) 138}.
%
\bibitem{HELAC1}
G.~Bevilacqua \textit{et al.},
  \href{https://doi.org/10.1016/j.cpc.2012.10.033}{\emph{Comput. Phys. Commun.} \textbf{184} (2013) 986}.
%
\bibitem{HELAC2}
G.~Bevilacqua, H.~B.~Hartanto, M.~Kraus, T.~Weber, M.~Worek,
  \href{https://doi.org/10.1007/JHEP03\%282020\%29154}{\emph{JHEP} \textbf{03} (2020) 154}.
%
\bibitem{Denner:2015yca}
A.~Denner and R.~Feger, 
  \href{https://doi.org/10.1007/JHEP11(2015)209}{\emph{JHEP} {\bfseries  11} (2015) 209}. 
%
\bibitem{Denner:2016wet}
A.~Denner, J.-N. Lang, M.~Pellen and S.~Uccirati, 
  \href{https://doi.org/10.1007/JHEP02(2017)053}{\emph{JHEP} {\bfseries 02} (2017) 053} 
%
\bibitem{Stremmer}
D.~Stremmer, M.~Worek,
\href{https://doi.org/10.1007/JHEP02\%282022\%29196}{\emph{JHEP} \textbf{02} (2022) 196}.
%
\bibitem{Demartin}
F.~Demartin, F.~Maltoni, K.~Mawatari, B.~Page, M.~Zaro,
  \href{ 	
https://doi.org/10.1140/epjc/s10052-014-3065-2}{\emph{Eur. Phys. J. C} \textbf{74} (2014) 3065}.
%

\end{thebibliography}
\end{document}